# Attack Tree Generation via Process Mining

Alyzia-Maria Konsta⊙, Gemma Di Federico⊙, Alberto Lluch Lafuente⊙, and Andrea Burattin⊙

DTU Compute, Technical University of Denmark, Kgs. Lyngby, Denmark
{akon,gdfe,albl,andbur}@dtu.dk

**Abstract.** Attack Trees are a graphical model of security used to study threat scenarios. While visually appealing and supported by solid theories and effective tools, one of their main drawbacks remains the amount of effort required by security experts to design them from scratch. This work aims at remedying this by providing a method for the automatic generation of Attack Trees from attack logs. The main original feature of our approach w.r.t. existing ones is the use of Process Mining algorithms to synthesize Attack Trees, which allow users to customize the way a set of logs are summarized as an Attack Tree, for example by discarding statistically irrelevant events. Our approach is supported by a prototype that, apart from the derivation and translation of the model, provides the user with an Attack Tree in the RisQFLan format, a tool used for quantitative risk modeling and analysis with Attack Trees. We use literature case studies to illustrate and explore the capabilities of our approach.

## 1 Introduction

The use of electronic devices has become an integral part of our daily life. These devices collect a huge amount of personal data, which are stored locally or on remote server systems. The increasing complexity of such systems has also incremented their vulnerability, making it critical to protect the data. Therefore, it is crucial to identify these weaknesses in order to improve the systems.

One way to detect and evaluate the threats of a system is through the use of graphical security models such as the Attack Trees [25]. Attack Trees are a graphical representation of the potential attacks that a system could receive. The tree-structured graphical representation provided by this framework places it in a clear and easy way to identify the weaknesses of the system.

Security experts work on building Attack Trees and use tools [4] to analyze the potential risks. A drawback of Attack Trees is that there is a gap [16] between research and the actual employment as the experts have to design these structures by hand, and the procedure can be tedious and error-prone (e.g., attacks can be wrongly modeled or over/underestimated). Instead, by exploiting the event log of a violated system, we can extract knowledge and precisely follow the attacker's steps. In fact, the log collects all the information needed to characterize the behaviors in an attack. Despite efforts from the research community, there is no tool to automatically derive Attack Trees from event data. Most tools that synthesize Attack Trees automatically are based on traces collected from existing models for the system under study (see [15] for a survey of such works).

The work proposed in this paper makes use of Process Mining techniques [2] to automatically derive an Attack Tree of a system from an event log.

The core of the proposal (cf. Figure 1) consists of the derivation of a Process Tree representing the behavior of the system, through the use of a Process Discovery algorithm. The obtained model is then translated into an Attack Tree.

The aim of discovering the process is to derive a model that faithfully describes the behavior of a system, by balancing the following dimensions. The model should *accurately* reproduce the cases recorded in the log, as well as *generalize* them so that it is able to reproduce future instances of the process, but at the same time not allow for *unobserved* behaviors. The model represents *dependencies* between events, but it should also be simple to be *understandable* by the experts. These criteria can be balanced as parameters of the process discovery algorithms. For example, the noise threshold of the Inductive Miner (one of the commonly used process mining algorithms) can be used to set focus on the discovery of either frequent or infrequent behaviors. For these reasons, the use of Process Mining makes it a valuable tool for the analysis of attacked systems. Indeed, the ability to customize and balance the discovery process constitutes one of the main original features of our work, w.r.t. existing works known from the literature [15].

Although Process Trees already hold relevant information, the security experts are not familiar with this structure, and it might be tedious for them to extract the important information. What is more, there are multiple tools [4] available that analyze different metrics of the Attack Trees to gain insightful knowledge about the security of the system. These kinds of tools cannot operate with Process Trees. Therefore, the necessity of translation. The work in this paper also supports the formalization and correctness of the translation. The approach has been implemented as a Python tool, and it is freely available[1]. The Attack Trees are exported in the RisQFLan format [4]: a tool that can be used for quantitative security risk modeling and analysis. The proposal constitutes a semi-automatic way of producing Attack Trees, as the tree can be subsequently modified and adapted.

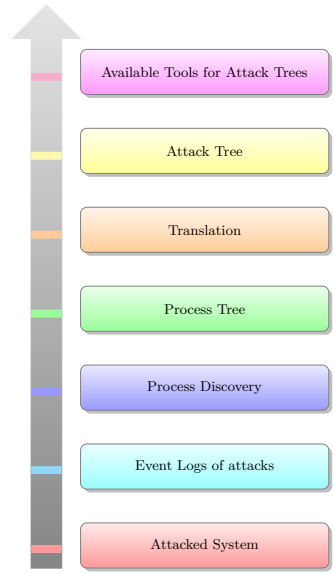

Fig. 1: Approach overview

Indeed, we envision our approach as closing the gap between threat modeling, assessment tools, and tools able to extract attack logs. To illustrate this, we use models for the literature of Attack Trees, as implemented in RisQFLan, in order to evaluate the capabilities of our approach when it comes to reconstructing an

---

[1] https://github.com/gemmadifederico/PTtoAT

Attack Tree. This is realized by generating attack traces for the same Attack Tree with different attacker profiles. Each attack profile yields a different set of traces and, ultimately, a different Attack Tree. We compare the obtained trees in terms of their similarity to the original tree.

In summary, the main contributions of this work are:

- A novel approach to obtain Attack Trees from logs of malicious activities (Section 3). The main differentiating feature of our approach is the use of process mining techniques, enabled by a novel transformation of Process Trees into Attack Trees.
- A prototype implementation[1] of our approach that can bridge the gap between security analyzers and threat modeling tools. We illustrate this in Section 4 with the state-of-the-art threat analysis tool (RisQFlan), that we use to evaluate the capability of our approach to reconstruct Attack Trees.

The rest of the paper includes a gentle introduction to Attack Trees and Process mining (Section 2), a discussion of related works (Section 5), and some concluding remarks (Section 6). We also include an Appendix with the formal proof of the correctness of our transformation, a discussion on how to handle loops in Process Trees, and tests for the correctness and scalability of our implementation.

## 2 Background

This section provides notions useful for the understanding of the paper. We introduce Attack Trees in Section 2.1. Then, Process Trees are presented (Section 2.2), along with a discussion of how to derive them automatically using process mining techniques (Section 2.3).

### 2.1 Attack Trees

In order to assess a system's security, Schneier proposed a technique called Attack Tree [25]. An Attack Tree is a graphical tree-structured representation of the system's security depicting possible attacks.

The tree structure of the graph highlights the vulnerabilities of the system and helps developers focus on the weak spots when they implement countermeasures [30]. The main idea behind the Attack Tree is to decompose the tasks of an attack into smaller tasks, thus making it easier to describe and quantify different metrics. When different attack tasks are connected to each other it means that there is a decomposition/refinement relationship but, the actual nature of the decomposition is expressed via operators. With the Attack Trees, one can capture multiple attacks derived from physical, technical, or even human vulnerabilities [30]. Since Schneier introduced the Attack Trees, multiple approaches and formal semantics have been proposed in the literature. In the scope of this work, we will try to include the most common operators used in the literature to cover most of the cases. To our knowledge, the most common operators are disjunction, conjunction [22], sequential conjunction [13], and the exclusive choice which was detected by Kordy et al. [16].

An example of an Attack Tree is shown in Figure 2. The main components of an Attack Tree can be narrowed down into three categories:

- **Root node:** the root node is the global goal of the attack. For example, in Figure 2 the global goal of the attacker is to "Rob Bob".
- **Children of a node:** are refinements of the parent's goal into sub-goals.
- **Leaf nodes:** these are basic attacks (i.e., tasks), that can not be further refined, that the attacker must perform in order to achieve their goal.

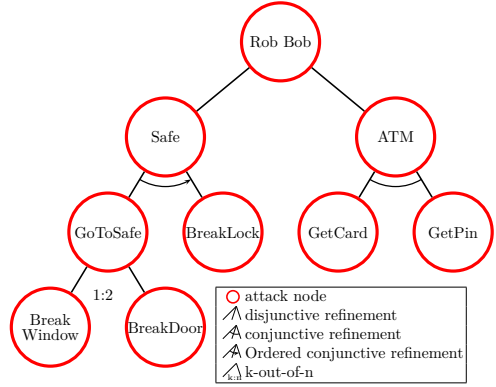

Fig. 2: Example Attack Tree

We can observe in Figure 2 that the refinements of an attack are represented in four different ways. For example the "Go to the safe" attack is refined into two attacks: "Break window" and "Break Door". Exclusively one of the subgoals should be fulfilled in order to achieve the parent goal. This kind of node is a *xor* node. We represent this kind of node as k-out-of-n, to align with the representation of ter Beek et al. [4]. The k-out-of-n indicates that k nodes should be achieved from n to fulfill the goal; in the *xor* case, exactly 1-out-of-n node should be achieved. For example, in our case, if the attacker has only one bomb for use, they will be able to break either the door or the window, but not both. Furthermore, the "(get money from) Safe" attack is refined into two sub-goals using the sequential conjunction [13]. In this case, all the children must be fulfilled in the given order – First the "Go to the safe" and then the "Break the lock". For the "ATM" attack a conjunction is introduced, which means that both the children must be achieved but the order is irrelevant. Finally, the "Rob Bob" attack is refined using a disjunction. In this case, one of the children must be fulfilled, but theoretically, the attacker can execute both. The difference with the exclusive operator is that in the *xor* case the attacker will only be able to fulfill one of the child nodes. It is worth mentioning that the *xor* operator is not broadly used in the literature on Attack Trees. In fact, we only came across one work mentioning the exclusive choice and it was referring to Fault Trees [16]. Fault Trees are also DAG-based structures and we decided to include this operator for completeness since one might be obligated to include such an operator in their analysis.

### 2.2  Process Trees

Process Trees are graph-based models that capture sound process models. These trees capture the relationships among activities of the process in a hierarchical fashion. Specifically, the inner nodes of a Process Tree represent the operators (dictating the order in which children should be executed) and the leaves represent the activities.

Five types of operators can be represented in a Process Tree, and they are the sequential operator ($\rightarrow$), the exclusive choice ($\times$), the parallel composition ($\wedge$), the redo loop ($\circlearrowleft$) and the inclusive choice, i.e. OR, ($\vee$). Let's consider the example in Figure 3. We can observe that the actions are on the leaves and the inner nodes are operators. The $\rightarrow$ operator indicates that its children should be executed in sequential order, meaning that the left sub-tree should be executed first. In the example, every process starts with executing activity $a$ followed by the sub-tree with the OR operator ($\vee$). After the OR, the sub-tree of the redo loop is executed, and finally the exclusive choice. The OR operator executes at least one of its children. The $\circlearrowleft$ redo loop operator has to have at least two children. The first child is in the "do" part (i.e., the part that has always to be executed) and the other children are "redo" parts. The redo loop starts the execution from the leftmost child and can loop back through any of its children. In the example, after the execution of the leftmost child of the redo loop, activity $f$ can be performed. The $\times$ operator is an exclusive choice, i.e. only one of the children has to execute. The last operator is the parallel $\wedge$, which can be observed in the example between the exclusive choice and the $d$ activity, and indicates that all the children have to be executed. The Process Tree in the example can be summarized textually as:

$$\rightarrow(a, \vee(b,c), \circlearrowleft(\rightarrow(\wedge(\times(i,j),d),e),f),\times(g,h))$$

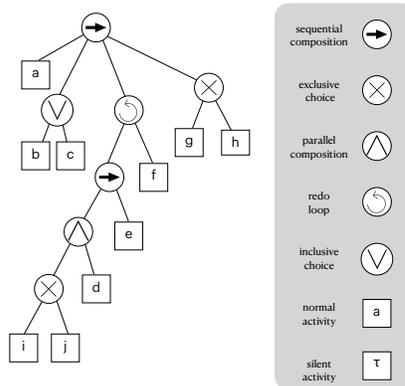

Fig. 3: Example of a Process Tree

The Process Tree notation also includes the $\tau$ activity, a silent activity that cannot be observed. The silent activity can be used in $\times(a,\tau)$ to indicate that activity $a$ can be skipped. More precisely, we follow the definitions introduced by van Zelst et al. [31] and Leemans [18] for the inclusive choice operator. In general, $\tau$ acts as the empty sequence $\epsilon$, that can be removed from sequences.

## 2.3 Process Mining

Process Mining consists of three techniques which are process discovery, conformance checking, and process enhancement [2].

In the context of this paper, we only focus on process discovery. Process discovery takes as input a log recording the execution of activities and produces a process model representing the process observed in the log. Several process discovery algorithms exist, such as the Heuristic Miner [29] and the Inductive Miner [20]. Algorithms can produce different types of process models, such as Petri Nets, Process Trees, Causal Nets, etc. Important quality criteria to select the discovery algorithm is the rediscoverability [19] of the process model. I.e., given an event log that contains information about the process, the algorithm is able to produce a model equivalent to the original process (up to some equivalence notion).

In order to apply Process Mining, it is necessary to have execution logs, recording which activities have been executed. These logs are called event logs. An example of an event log is reported in Table 1. Event logs are grouped into cases, that are process instances. Each case consists of events, that correspond to activities or tasks performed by a process participant. Each event has an associated set of attributes, such as the activity name, the timestamp of the event, or the resource that executed the activity. For example, the first event in Table 1 refers to the execution of the activity *register request*, on the date 30/12/2010 at 11.02, by a user called Pete. The set of attributes of an event log can be extended. For the scope of this work, the minimum information required is the case id and the activity attribute. To position the event log in the context of this paper, a log collects information regarding attacks. The event log can both represent successful and unsuccessful attempts. For example, in Figure 2 the attacker might try to break the window multiple times. This is represented as different events in the event log. However, in the Attack Tree, as shown in Figure 2, this action is represented once by a label.

| CaseID | Properties | | |
| --- | --- | --- | --- |
| | Timestamp | Activity | Resource |
| 1 | 30-12-2010:11.02 | register request | Pete |
| | 31-12-2010:10.06 | examine thoroughly | Sue |
| | 05-01-2011:15.12 | check ticket | Mike |
| | 07-01-2011:14.24 | reject request | Pete |
| 2 | 30-12-2010:11.32 | register request | Mike |
| | 30-12-2010:12.12 | check ticket | Mike |
| | 30-12-2010:14.16 | examine casually | Pete |

Table 1: Example of an Event Log in [27]

An event log can be processed by a process discovery algorithm in order to derive a process model. A discovery algorithm that ensures the rediscoverability property is the Inductive Miner [20] (IM). The Inductive Miner algorithm makes use of a divide-and-conquer approach to decompose the event log into smaller sublogs, in order to construct a Process Tree. The algorithm separates activities, selects an operator, and splits the log, after that it iterates over the sublogs until convergence. The IM provides the guarantee that it can re-discover the process model from an event log since it relies on the directly following relation between all pairs of activities in the event log. For a detailed explanation of the algorithm, please refer to the book by van der Aalst [27]. The IM returns a Process Tree. Since the objective of this work is to represent attacks, we want to ensure that our model is sound, meaning that every activity can participate in a process instance and it is ensured that the process always terminates properly. For this reason, we decided to consider the Inductive Miner as a valid solution to derive an Attack Tree.

## 3   From Process Trees To Attack Trees

In this section, we carefully explain the details of our approach, with a particular focus on the transformation from a Process Tree into an Attack Tree. What is more, we formalize and prove the correctness of the translation.

The approach is divided into two phases: the mining and the translation. In the mining phase, the behavior of an attacker, collected in the event log, is represented in the form of a process model. The model captures activities and their dependencies in a conceptual model, i.e. a Process Tree. The Process Tree is derived by the use of the Inductive Miner algorithm. As previously mentioned, through fine-tuning the algorithm's parameters, it is possible to emphasize particular viewpoints e.g., take into consideration infrequent behaviors or deal with incomplete event logs. The resulting Process Tree is translated into the corresponding Attack Tree in the second phase of the approach. The Attack Tree is also converted in the RisQFLan format so that it can be used in the tool for further analysis. Before going into detail with the transformation, we introduce the semantics of both modeling languages.

In particular, we start providing trace-based semantics of Attack Trees (Section 3.1) and Process trees (Section 3.2), to support the formalization and correctness of our translation. Then, the main idea of the transformation is provided in Section 3.3 and the specific transformation rules are provided in Section 3.4. We prove that the source Process Tree can produce the same potential traces as the translated Attack Tree (in the Appendix C). The case of the loop operator is reported in the Appendix B.

### 3.1   Attack Tree Semantics

In our semantic definitions, we assume there is an alphabet $A$ of symbols representing actions. We use standard operations and notations for traces, including a trace interleaving function $\| : (A^* \times A^*) \to (A \cup A)^*$ defined as follows:

$$\emptyset \| A_1 = A_1 \quad (\{w\} \cup A_1) \| A_2 = (w \| A_2) \cup (A_1 \| A_2)$$
$$A_1 \| \emptyset = A_1 \quad \alpha w_1 \| \beta w_2 = (\alpha(w_1 \| \beta w_2)) \cup (\beta(\alpha w_1 \| w_2))$$
$$\epsilon \| w = \{w\} \quad w \| \epsilon = \{w\}$$

We sometimes denote concatenation of traces by juxtaposition and sometimes we explicitly use a concatenation operator $\cdot$ when it improves readability. The operator $\cdot$ is lifted to sets of traces as usual (pairwise concatenation).

The formal syntax and semantics of Attack Trees are given in Table 2. Attack Trees are terms generated by $T$ in the grammar. The trace-based semantics of Attack Trees are given by function $[\![\ ]\!]_t : T \to A^*$, which maps each tree into a

$$T := \alpha \mid and(\alpha, T_1, \ldots, T_n) \mid or(\alpha, T_1, \ldots, T_n) \mid xor(\alpha, T_1, \ldots, T_n) \mid sand(\alpha, T_1, \ldots, T_n)$$

where $\alpha \in A$ and $n \leq 1$

$$[\![\alpha]\!]_t = \{\alpha\}$$

$$[\![and(\alpha, T_1, \ldots, T_n)]\!]_t = ([\![T_1]\!]_t \| \ldots \| [\![T_n]\!]_t) \cdot \alpha$$

$$[\![or(\alpha, T_1, \ldots, T_2)]\!]_t = \bigcup_{i \in \{0, \ldots, n\}} \{([\![T_i]\!]_t \| w) \cdot \alpha \mid \exists w'.w \cdot w' \in \|_{j \in \{0, \ldots, n\} \setminus \{i\}} [\![T_j]\!]_t\}$$

$$[\![xor(\alpha, T_1, \ldots, T_n)]\!]_t = ([\![T_1]\!]_t \cdot \alpha) \cup \cdots \cup ([\![T_n]\!]_t \cdot \alpha)$$

$$[\![sand(\alpha, T_1, \ldots, T_n)]\!]_t = [\![T_1]\!]_t \cdot \cdots \cdot [\![T_n]\!]_t \cdot \alpha$$

Table 2: Attack Tree Syntax and Semantics

set of action sequences. Our Attack Tree semantics are based on the semantics of Attack Trees supported by RisQFlan [4] since our purpose is to be able to produce Attack Trees for said tool. It is worth to say that, as observed by other authors (see discussion in [21]) there is no common agreement on the meaning of Attack Trees. The semantics presented here, in addition to being compatible with RisQFlan, are close to the ones presented in [21] with some minor differences discussed in said paper (e.g. labels in inner nodes).

Table 2 defines the function by providing rules for creating the traces for every operator.

As can be seen from Table 2, an Attack Tree can be a single node $\alpha \in A$, where $A$ is a set of actions or a combination of $n$ subtrees with a new node. The $and(\alpha, T_1, \ldots, T_n)$ operator, denotes a conjunction where node $\alpha$ is the parent of the $n$ subtrees $T_1$ to $T_n$. In this case, in order to reach node $\alpha$ all of the $n$ subtrees $T_1$ to $T_n$ should be achieved. Accordingly, the $or(\alpha, T_1, \ldots, T_n)$ operator, depicts a disjunction where node $\alpha$ is the parent of the $n$ subtrees $T_1$ to $T_n$. In this case, in order to reach node $\alpha$ at least one of the $n$ subtrees $T_1$ to $T_n$ should be achieved. The $xor(\alpha, T_1, \ldots, T_n)$ depicts an exclusive or, meaning that in order to achieve the parent goal, exactly one of the subtrees should be fulfilled.[2] Finally, the $sand(\alpha, T_1, \ldots, T_n)$ introduces the sequential conjunction meaning that the events are ordered [13]. We chose to include the most commonly used operators we came across in the literature. One can decide to use a subset of the aforementioned operator according to their purpose.

---

[2] Readers familiar with [21] may recognize that this corresponds to the semantics of disjunction on said paper. Consequently, our approach can be easily adapted to produce Attack Trees in the format of tools that follow [21]

### 3.2   Process Tree Semantics

A Process Tree represents activities in a hierarchical order. The inner nodes of a Process Tree are operators and the leaves are activities, $\alpha \in A$. We formalize the syntax and semantics of Process Trees similarly to Attack Trees. Table 3 provides the grammar for Process Trees (terms generated by $P$) as well as the semantic function $[\![\,]\!]_p : P \to A^*$, which maps each Process Tree into a set of traces, following the standard meaning of Process Trees [2].

---

$$P := \alpha \mid and(P_1, \ldots, P_n) \mid or(P_1, \ldots, P_n) \mid xor(P_1, \ldots, P_n) \mid \to (P_1, \ldots, P_n) \mid \circlearrowleft (P_1, \ldots, P_n)$$

$$\text{where } \alpha \in A \text{ and } n \leq 1$$

---

$$[\![\alpha]\!]_p = \{\alpha\}$$

$$[\![\tau]\!]_p = \{\}$$

$$[\![and(P_1, \ldots, P_n)]\!]_p = ([\![P_1]\!]_p \| \ldots \| [\![P_n]\!]_p)$$

$$[\![or(P_1, \ldots, P_n)]\!]_p = \bigcup_{i \in \{0, \ldots, n\}} \{([\![P_i]\!]_p \| w) \mid \exists w', w \cdot w' \in \|_{j \in \{0, \ldots, n\} \setminus \{i\}} [\![P_j]\!]_p\}$$

$$[\![xor(P_1, \ldots, P_n)]\!]_p = [\![P_1]\!]_p \cup \cdots \cup [\![P_n]\!]_p$$

$$[\![\to(P_1, \ldots, P_n)]\!]_p = [\![P_1]\!]_p \cdot \cdots \cdot [\![P_n]\!]_p$$

$$[\![\circlearrowleft (P_1, \ldots, P_n)]\!]_p = [\![P_1]\!]_p \cdot (([\![P_2]\!]_p \cup \cdots \cup [\![P_n]\!]_p) \cdot [\![P_1]\!]_p)^*$$

---

Table 3: Process Tree Syntax and Semantics

According to Table 3 a Process Tree can be a single node or a combination of an operator and $n$ subtrees. The $and(P_1, \ldots, P_n)$ operator defines a conjunction between $n$ subtrees, which means that all the subtrees $P_1$ to $P_n$ should be executed in order to reach the goal. The sequence operator $\to (P_1, \ldots, P_n)$, defines a sequential relation between the subtrees $P_1$ to $P_n$, which means that $P_1$ should be executed before $P_2$ and so on. We can perceive this operator as a parental relationship between $P_{n-1}$ and $P_n$, where $P_n$ is the parent node. We are also defining the $xor(P_1, \ldots, P_n)$ which depicts the exclusive choice among the $n$ subtrees $P_1$ to $P_n$. Finally, we introduce the $or(P_1, \ldots, P_n)$ operator, which defines a disjunction, where at least one of the $n$ subtrees should be executed in order to reach the goal. Finally, we define the redo loop $\circlearrowleft$ operator, where the leftmost child is always executed and can loop back through any of the other children and execute the first child again. The repetition is not mandatory, that is why we enclose the traces with the Kleene star, indicating a possible repetition.

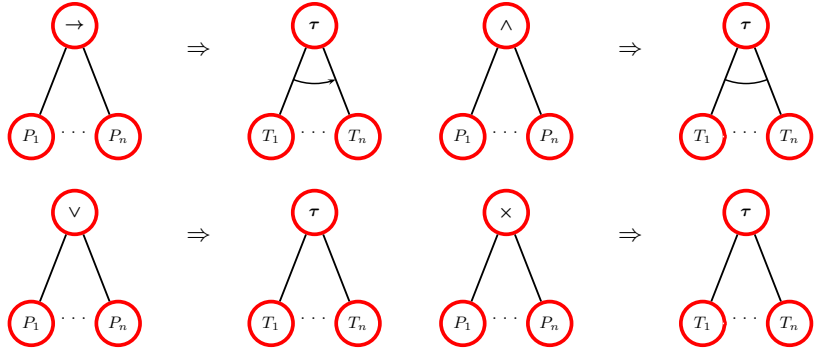

Fig. 4: Transforming Process Trees into Attack Trees, formally

$$\mathsf{p2t}(a) = a \tag{1}$$
$$\mathsf{p2t}(\rightarrow (P_1, \ldots, P_n)) = sand(\tau, \mathsf{p2a}(P_1), \ldots, \mathsf{p2t}(P_n)) \tag{2}$$
$$\mathsf{p2t}(and(P_1, \ldots, P_n)) = and(\tau, \mathsf{p2t}(P_1), \ldots, \mathsf{p2t}(P_n)) \tag{3}$$
$$\mathsf{p2t}(or(P_1, \ldots, P_n)) = or(\tau, \mathsf{p2t}(P_1), \ldots, \mathsf{p2t}(P_n)) \tag{4}$$
$$\mathsf{p2t}(xor(P_1, \ldots, P_n)) = xor(\tau, \mathsf{p2t}(P_1), \ldots, \mathsf{p2t}(P_n)) \tag{5}$$

Table 4: Transforming Process Trees into Attack Trees, formally

### 3.3   Basics of the Transformations

Attack Trees, by definition, are composed of a main goal (i.e. the root node), sub-goals (intermediate nodes), and actions (leaves). The main goal is decomposed into sub-goals. Each attack consists of components required to perform the attack. On the other side, Process Trees belong to the family of process models. A process model describes the flow of activities that are executed in order to accomplish a specific goal. The goal of a process model is to describe activities and relationships, and their execution order [2]. In a Process Tree, internal vertices represent operators, and leaves represent activities. The main difference between the two languages is that an Attack Tree explicitly models activities and goals, while a Process Tree does not directly model the goal, since it is the objective of the model representation itself. To translate the concept of goal into the Process Tree, we consider the root node (from the Attack Tree) as the last activity to be executed in a Process Tree. More specifically, rephrasing it under a process perspective, to achieve a goal you first need to execute the list of activities required and, in the end, you reach the goal node. Hence, the last activity executed in the Process Tree replaces the root node in the Attack Tree.

However, since the concept of goal is not embedded in the Process Tree, we had to introduce the notion of observable and non-observable actions: an action $\alpha \in A$ collected by the information system, which contributes to the achievement of the goal is called ***observable***, while an action $\alpha \in A$ that cannot be directly mapped to an execution of an activity is called ***non-observable***. The distinction between observable and non-observable actions is necessary since Attack Trees have a goal-oriented and self-explainable structure, while Process Trees, in order to be executed, require that each activity node is observable in the event log. We hence denote all non-observable actions by $\tau, \tau_i$, which is commonly used to denote silent actions [1]. These will serve in the Attack Tree as intermediate nodes in some cases, as we shall see.

As introduced in Section 2, the Attack includes four operators, conjunction (*and*), disjunction (*or*), exclusive choice (*xor*), and sequential conjunction (*sand*). On the other side, Process Tree defines four different operators that are the sequential ($\rightarrow$), the exclusive choice (*xor*), the parallel composition (*and*), the disjunction (*or*), and the redo loop ($\circlearrowleft$). In order to translate an Attack Tree into a Process Tree, the latter must be able to represent the operators of the former. We can see a correspondence between the operators of the two trees. The conjunction already finds a definition in the Process Tree, that is the parallel operator (*and*), the disjunction (*or*), and the exclusive choice (*xor*) can be found in both trees and finally, the sequential conjunction (*sand*) can be paired with the sequential operator ($\rightarrow$). The main subtle difference we need to take into account is that Attack Trees have action-labeled internal nodes. Our transformation takes care of this introducing non-observable silent actions $\tau$.

The redo loop ($\circlearrowleft$) operator is more complicated, as there is not a 1 to 1 mapping in the Attack Tree. The redo loop, states that the leftmost branch is always executed, and can loop back through any of its other children and then execute the leftmost child again. We can see the traces produced in Table 3, where the traces of the leftmost child are always produced. However, the traces of the loop are denoted by the Kleen star *, meaning that the procedure might be repeated or not. Handling the loop operator is out of the scope of this paper, but we include a discussion in the Appendix B.

### 3.4   Transformation Rules

We are now ready to present one of the main contributions of our paper, namely the transformation of Process Trees into Attack Trees.

The transformation is formally provided by the function $\mathtt{p2t} : P \rightarrow T$, which transforms Process Trees into an Attack Tree. The function is formally defined in Table 4. The definition is by structural induction and the four recursive cases are graphically depicted in Figure 4. The top-left rule in Figure 4 concerns the sequence construct. As observed in Figure 4 the sequence operator ($\rightarrow$) can be replaced by the sequential conjunction (*sand*) operator. Both operators define an order between the nodes involved. The top-right rule in Figure 4 represents the translation of the $\wedge$ (*and*) operator of the Process Tree into conjunction, the *and* operator of the Attack Tree. The bottom-left rule in Figure 4 represents the translation of the $\vee$ (*or*) operator of the Process Tree into disjunction, the *or* operator of the Attack Tree. The bottom-right rule in Figure 4 represents the translation of the × (*xor*) operator of the Process Tree into exclusive or the *xor* operator of the Attack Tree. Common to all four rules is the fact that a non-observable action $\tau$ is introduced as the root of each sub-tree.

One of the main results of the paper is that our translation is correct, as stated in the following theorem, whose proof can be found in the Appendix A.

**Theorem 1.** *Let $P$ be a Process Tree. Then $[\![P]\!]_p = [\![\mathsf{p2t}(P)]\!]_t$.*

## 4   Reconstructing Attack Trees generated by RisQflan

We address in this section the following research question: Can our approach reconstruct an Attack Tree based just on observing attack traces based on said Attack Tree?

To address the question, we use a model of the literature of Attack Trees, as implemented in RisQFlan, namely the Bypassing 802.1x Attack Tree [6]. Said Attack Tree is provided as part of the RisQflan tool as an example.

The Attack Tree can be seen in Figure 5. The nodes represent the following attacker actions and (sub)goals (as presented in [6]): *"A - Bypass 802.1x, B - Hijack 802.1x Authenticated Session, C - Disconnect Client, D - Find 802.1x Authenticated Victim, E - MiM 802.1x Session, F - Find Unauthenticated Victim, a - Eavesdrop on 802.1x Authenticated Client, b - Use MAC Address of AP, c - Send MAC Disassociate, d - Impersonate 802.1x Authenticated Client, e - Eavesdrop on New Unauthenticated Client, f - Impersonate AP, g - Impersonate New Unauthenticated Client"*. The node K is sub-goal introduced in [26] for technical reasons.

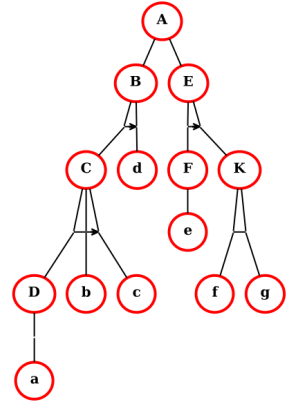

Fig. 5: Bypassing Attack Tree

To test the reconstruction capabilities of our approach, we will first generate traces with RisQFLan for the given Attack Tree and then use them to generate an Attack Tree, hopefully similar to the original one.

In the RisQflan tool, one can specify an attacker profile that depicts the behavior of a specific attacker. The attacker profiles are represented as probabilistic automata describing the possible attack steps and their probabilities/rates. We have used 4 different attacker profiles, which are provided in the RisQFlan model:

- Best: An attacker aware of the optimal order of the attacks in order to achieve the goal [4]. We are introducing two different Best attackers. The first one (called Best) is executing the left branch of the OR operator under the root in the Bypassing tree and the second one (called BestB) the right one.
- Average: An attacker randomly trying attacks until achieving the main attack goal or a wrong order leads to failure. We filter the logs to include only successful attacks since we are not interested in the failed ones [4].

– Worst: Like Average but chooses attacks with a probability inversely proportional to the order used by Best [4].

For each one of the above attacker profiles using the RisQflan tool, we generate different log files, with possible traces generated by each one of the attackers. We then generate the corresponding .xes log file containing only the successful attacks, which serves as input to our code that translates a process tree into an Attack Tree. The process tree is obtained by using a generation function included in the Process Mining library PM4py[3], and specifically the function *pm4py.discover_process_tree_inductive*. The function takes as parameters the log file and the noise threshold, which is set by default at 0.

The obtained Attack Trees are shown in Figures 6–9. We can notice that based on the attacker profile we get completely different Attack Trees depending on the behavior of the attacker (since they generate different sets of traces). The Best and BestB attackers know exactly the steps that should be followed to succeed in the attack, so the generated traces produce an Attack Tree with only one $SAND$ node, representing the sequence in the actions should be executed. None of them produce the departing Attack Tree, but they faithfully summarize the two possible attackes: the Attack Tree in Figure 6 depicts the left branch of the initial Bypassing Attack Tree in Figure 5, whereas the Attack Tree in Figure 7 represents the right one.

The Average attacker is trying out randomly all the actions, which results in the Attack Tree on Figure 8. We can see that this Attack Tree differs a lot from the original Attack Tree. This is happening because the attacker is randomly executing actions. One example found in the log file is the following trace: ['b', 'd', 'c', 'e', 'a', 'F', 'f', 'g', 'K', 'D', 'E', 'A']. We can see that the right branch of the Attack Tree in Figure 5 is executed correctly ['e', 'F', 'f', 'g', 'K', 'E', 'A'], but the noisy actions happening in between the correct trace (which could be interpreted as blind and useless attempts of the attacker) cannot be filtered out with process mining. The same phenomena is also happening in Figure 9 for the Worst attacker.

The conclusion of our experiments is that the rationality and effectiveness of attacks (and their logs) can have a significant impact in the quality of the obtained Attack Trees. Random or uninformed attackers may produce logs with enough noise to make the obtained Attack Trees hardly usable. This opens a further research question of how to filter out noise from attack logs for the benefit of mining understandable and meaningful Attack Trees.

## 5   Related Work

To the best of our knowledge, there are no other works exploiting the logs of violated systems leveraging Process Mining to derive the corresponding Attack Tree. Thus we present some works considering the Automatic Generation of Attack Trees and point out the advantages of our work compared to the existing ones. For a more detailed study of the current approaches for the automatic generation of Attack Trees, we encourage the reader to refer to [15].

---

[3] `https://pm4py.fit.fraunhofer.de`

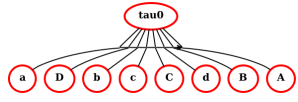

Fig. 6: Best attacker

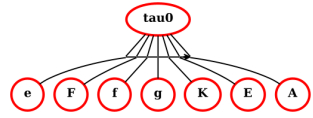

Fig. 7: BestB attacker

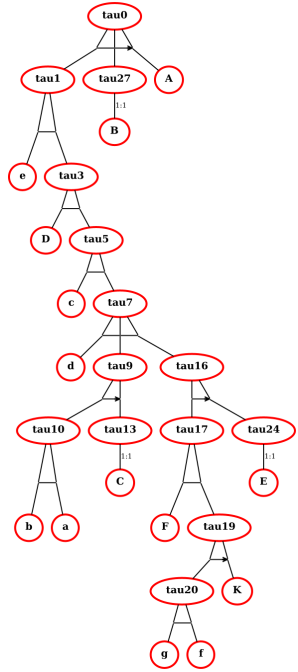

Fig. 8: Average attacker

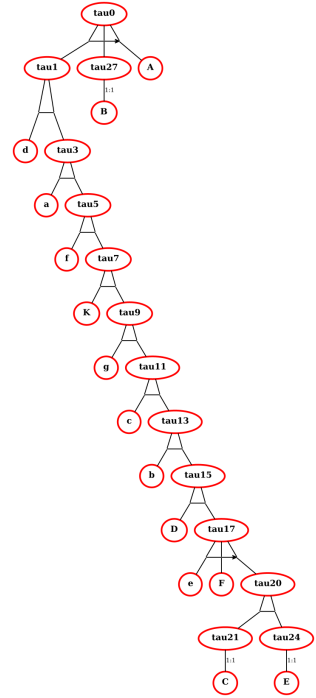

Fig. 9: Worst attacker

According to [15], there are only 3 works that generate Attack Trees from a security analysis of a system model, namely [12, 23, 28]. The main idea in such approaches is to use a model of the system under study, specify security properties, and use counterexamples for such properties as attack traces that are then summarized in Attack Trees. The main difference of our approach is that we use process mining algorithms, which allow us to accommodate the user preference in balancing features such as precision and inclusion of statistically (ir)relevant events.

Another family of approaches is the so-called *vulnerability-driven* approaches, which use libraries of predefined Attack Tree templates as the starting point. According to [15], there are only 5 works that generate Attack Trees using this approach, namely [5, 11, 14, 17, 24]. The main difference with our work is that they often do not depart from attack traces and often necessitate domain experts to instantiate the templates for the system under study. An exception is the work presented in [24] which provides an Attack Tree for a single attack log by parsing it given a suitable grammar of common patterns of attack (de)composition.

On a different line of research, process mining has been used in RisQFlan to refine attacker profiles from logs produced from simulations generated by the statistical analysis of RisQFlan models [7–10]. Attacker profiles are specific to RisQFlan's approach to threat modelling, and in this work we have focused on the more common approach to model only the declarative part of the threat model as an attack tree only. One may want to explore if the approach of [7–10] and our proposal can be combined to discover threat models enriching Attack Trees with additional descriptions.

## 6   Conclusion and Future Work

We have presented a tool-supported approach to derive Attack Trees directly from observed malicious activities in the form of logs. Our automatic derivation is intended to complement the manual design of the Attack Trees. The obtained trees can be seen as initial proposals that can then be adapted by domain experts. The paper proves that the translation not only is feasible (i.e., it is implemented and has thoroughly been tested) but it is also formally correct. The actual implementation targets the RisQFlan tool. Our tool can be adapted and integrated to connect with other Attack Tree tools as well or to import traces from other security analyzers as those in [15]. The empirical evaluation conducted in this paper has shown that the nature of the attack traces can have a significant impact in the quality of the obtained Attack Trees. Further investigations and research would be needed to address the problem of mitigating the effect of noise in logs to obtain more informative and useful Attack Trees, possibly enjoying properties such as admissibility, consistency and soundness [3]. As additional future work, we would like to extend the set of operators provided and the set of tools for which we produce an output. A related question is whether and how Process Mining techniques can be applied to synthesize Attack Defense Trees, an extension of Attack Trees that includes attacker and defender behavior.

## A    Proof of the Transformation Theorem

**Theorem 1.** *Let $P$ be a Process Tree. Then $[\![P]\!]_p = [\![\mathsf{p2t}(P)]\!]_t$.*

*Proof.* The proof is by structural induction, examining each case from Table 4 (*Transforming Process Trees into Attack Trees, formally* from the original paper), separately.

The base case is when $P = a$. This case is trivial. Indeed

$$
\begin{aligned}
& [\![\mathsf{p2t}(a)]\!]_t \\
= {} & [\![a]\!]_t && \text{(by definition of } \mathsf{p2t}) \\
= {} & \{a\} && \text{(by definition of } [\![\cdot]\!]_t) \\
= {} & [\![a]\!]_p && \text{(by definition of } [\![\cdot]\!]_p)
\end{aligned}
$$

Case $P = \to (P_1, \dots P_n)$. We will assume that Theorem holds for $P_1, \dots, P_n$, i.e. that $[\![P_i]\!]_p = [\![\mathsf{p2t}(P_i)]\!]_t$, $\forall i \in [1, \dots, n]$.

$$
\begin{aligned}
& [\![\to (P_1, \dots P_n)]\!]_p \\
= {} & [\![P_1]\!]_p \cdots \cdot [\![P_n]\!]_p && \text{(by definition of } [\![\cdot]\!]_p) \\
= {} & [\![P_1]\!]_p \cdots \cdot [\![P_n]\!]_p \cdot \tau && \text{(since } \tau \text{ is neutral)} \\
= {} & [\![\mathsf{p2t}(P_1)]\!]_t \cdots \cdot [\![\mathsf{p2t}(P_n)]\!]_t \cdot \tau && \text{(by induction)} \\
= {} & [\![sand(\tau, \mathsf{p2t}(P_1), \dots \mathsf{p2a}(P_n))]\!]_t && \text{(by definition of } [\![\cdot]\!]_t) \\
= {} & [\![\mathsf{p2t}(\to (P_1, \dots, P_n))]\!]_t && \text{(by definition of } \mathsf{p2t})
\end{aligned}
$$

Case $P = and(P_1, \dots P_n)$. We will assume that Theorem holds for $P_1, \dots, P_n$, i.e. that $[\![P_i]\!]_p = [\![\mathsf{p2t}(P_i)]\!]_t$, $\forall i \in [1, \dots, n]$.

$$
\begin{aligned}
& [\![and(P_1, \dots P_n)]\!]_p \\
= {} & [\![P_1]\!]_p \| \dots \| [\![P_n]\!]_p && \text{(by definition of } [\![\cdot]\!]_p) \\
= {} & [\![P_1]\!]_p \| \dots \| [\![P_n]\!]_p \| \tau && \text{(since } \tau \text{ is neutral)} \\
= {} & [\![\mathsf{p2a}(P_1)]\!]_t \| \dots \| [\![\mathsf{p2t}(P_n)]\!]_t \| \tau && \text{(by induction)} \\
= {} & [\![and(\tau, \mathsf{p2a}(P_1), \dots, \mathsf{p2a}(P_n))]\!]_t && \text{(by definition of } [\![\cdot]\!]_t) \\
= {} & [\![\mathsf{p2t}(and(P_1, \dots, P_n))]\!]_t && \text{(by definition of } \mathsf{p2t})
\end{aligned}
$$

Case $P = or(P_1, \dots P_n)$. We will assume that Theorem holds for $P_1, \dots, P_n$, i.e. that $[\![P_i]\!]_p = [\![\mathsf{p2t}(P_i)]\!]_t$, $\forall i \in [1, \dots, n]$.

$$
\begin{aligned}
& [\![or(P_1, \dots P_n)]\!]_p \\
= {} & \bigcup_{i \in \{0, \dots, n\}} \{([\![P_i]\!]_p \| w) \mid \exists w', w \cdot w' \in \|_{j \in \{0, \dots, n\} \setminus \{i\}} [\![P_j]\!]_p\} && \text{(by definition of } [\![\cdot]\!]_p) \\
= {} & \bigcup_{i \in \{0, \dots, n\}} \{([\![P_i]\!]_p \| w) \mid \exists w', w \cdot w' \in \|_{j \in \{0, \dots, n\} \setminus \{i\}} [\![P_j]\!]_p\} \cdot \tau && \text{(since } \tau \text{ is neutral)} \\
= {} & \bigcup_{i \in \{0, \dots, n\}} \{([\![\mathsf{p2a}(P_i)]\!]_t \| w) \mid \exists w', w \cdot w' \in \|_{j \in \{0, \dots, n\} \setminus \{i\}} [\![\mathsf{p2a}(P_j)]\!]_t\} \cdot \tau && \text{(by induction)} \\
= {} & [\![or(\tau, \mathsf{p2a}(P_1), \dots, \mathsf{p2a}(P_n))]\!]_t && \text{(by definition of } [\![\cdot]\!]_t) \\
= {} & [\![\mathsf{p2t}(or(P_1, \dots, P_n))]\!]_t && \text{(by definition of } \mathsf{p2t})
\end{aligned}
$$

Case $P = xor(P_1, \dots P_n)$. We will assume that Theorem holds for $P_1, \dots, P_n$, i.e. that $[\![P_i]\!]_p = [\![\mathsf{p2t}(P_i)]\!]_t$, $\forall i \in [1, \dots, n]$.

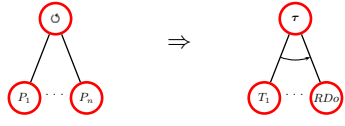

Fig. 10: Translating the $\circlearrowleft$ operator.

$$\begin{aligned}
&\llbracket xor(P_1, \ldots P_n) \rrbracket_p \\
&= \llbracket P_1 \rrbracket_p \cup \cdots \cup \llbracket P_n \rrbracket_p && \text{(by definition of } \llbracket \cdot \rrbracket_p) \\
&= \llbracket P_1 \rrbracket_p \cup \cdots \cup \llbracket P_n \rrbracket_p \cup \tau && \text{(since } \tau \text{ is neutral)} \\
&= \llbracket \mathsf{p2t}(P_1) \rrbracket_t \cup \cdots \cup \llbracket \mathsf{p2t}(P_n) \rrbracket_t \cup \tau && \text{(by induction)} \\
&= \llbracket xor(\tau, \mathsf{p2t}(P_1), \ldots, \mathsf{p2t}(P_n)) \rrbracket_t && \text{(by definition of } \llbracket \cdot \rrbracket_t) \\
&= \llbracket \mathsf{p2t}(xor(P_1, \ldots, P_n)) \rrbracket_t && \text{(by definition of } \mathsf{p2t})
\end{aligned}$$

## B  On loops

We discuss here some ideas for handling the loop operator of the Process Trees.

To tackle Process Trees with loops we will extend the formal syntax and semantics of Attack Tree and Process Tree semantics to also represent the notion of repetition. For Attack Trees this constitutes a novel operator not present in existing forms of Attack Trees, as far as we know.

The extension of Attack Trees with repetition consists of a new syntactic operator $\_^r$, whose semantics is the same as the Kleene star, i.e. $\llbracket T_1^r \rrbracket = \llbracket T_1 \rrbracket^*$. We use a different symbol $r$ to avoid confusion.

Similarly, we extend the syntax and semantics of Process Trees introducing the same kind of operator for repetitions. It is worth to note that the semantics of such operator is the Kleen star and hence different from the redo operator. Actually, the redo operator can be encoded with the new repetition operator as formalized by the following Lemma.

**Lemma 1.** *Let $P_1, \ldots, P_n$ be $n$ Process Trees, combined under the redo loop operator $P = \circlearrowleft (P_1, \ldots, P_n)$. Then $P = \circlearrowleft (P_1, \ldots, P_n)$ is equivalent to $P' = \rightarrow (P_1, \rightarrow (xor(P_2, \ldots, P_n), P_1)^r)$.*

*Proof.*

$$\begin{aligned}
&\llbracket \circlearrowleft (P_1, \ldots P_n) \rrbracket_p \\
&= \llbracket P_1 \rrbracket_p \cdot ((\llbracket P_2 \rrbracket_p \cup \cdots \cup \llbracket P_n \rrbracket_p) \cdot \llbracket P_1 \rrbracket_p)^* \\
&= \llbracket P_1 \rrbracket_p \cdot ((\llbracket xor(P_2, \ldots, P_n) \rrbracket_p) \cdot \llbracket P_1 \rrbracket_p)^* \\
&= \llbracket P_1 \rrbracket_p \cdot (\llbracket \rightarrow (xor(P_2, \ldots, P_n), P_1)^r \rrbracket_p) \\
&= \llbracket \rightarrow (P_1, \rightarrow (xor(P_2, \ldots, P_n), P_1)^r) \rrbracket_p
\end{aligned}$$

Now, the transformation function $\mathsf{p2t}$ is extended to cover the two new cases:

$$\mathsf{p2t}(P_1^r) = \mathsf{p2t}(P_1)^r$$

$$\mathsf{p2t}(\circlearrowleft (P_1, \ldots, P_n)^r) = sand(\tau, \mathsf{p2t}(P_1), \mathsf{p2t}(sand(\tau, xor(\tau, \mathsf{p2t}(P_2), \ldots, \mathsf{p2t}(P_n)), \mathsf{p2t}(P_1))^r)$$

The main subtle detail is that the transformation of the redo operator exploits the observation of Lemma 1. On Figure 10 we can see the translation of the redo loop ↻ operator of the Process Tree into the equivalent branch of the Attack Tree. Following the idea of Lemma 1, the final branch of the Attack Tree is a *sand*. The root of the *sand* is a $\tau$, the first child $T_1$ is $\mathsf{p2t}(P_1)$, and the second one is represented by the *RDo*, which hides the following branch of the tree, $sand(\tau, xor(\tau, \mathsf{p2t}(P_2)), \ldots, \mathsf{p2t}(P_n)), \mathsf{p2t}(P_1))^r$.

We can now extend Theorem 1 to cover Process Trees with redo loops.

**Theorem 2.** *Let $P$ be a Process Tree. Then $[\![P]\!]_p = [\![\mathsf{p2t}(P)]\!]_t$.*

*Proof.* The proof extended the cases of the proof of Theorem 1.

The case when $P = P_1^r$, where for induction we assume $[\![\mathsf{p2t}(P_1)]\!]_t = [\![P_1]\!]_p$:

$$
\begin{aligned}
&[\![\mathsf{p2t}(P_1^r)]\!]_t \\
&= [\![\mathsf{p2t}(P_1)^r]\!]_t \quad \text{(by definition of } \mathsf{p2t}) \\
&= [\![\mathsf{p2t}(P_1)]\!]_t^* \quad \text{(by semantics of } [\![\cdot]\!]_t) \\
&= [\![P_1]\!]_p^* \quad\quad\ \ \text{(by induction)} \\
&= [\![P_1^r]\!]_p \quad\quad\ \text{(by semantics of } [\![\cdot]\!]_p)
\end{aligned}
$$

Case $P = \circlearrowright (P_1, \ldots P_n)$. We will assume that Theorem holds for $P_1, \ldots, P_n$, i.e. that $[\![P_i]\!]_p = [\![\mathsf{p2t}(P_i)]\!]_t, \forall i \in [1, \ldots, n]$.

$$
\begin{aligned}
&[\![\circlearrowright (P_1, \ldots P_n)]\!]_p \\
&= [\![P_1]\!]_p \cdot (([\![P_2]\!]_p \cup \cdots \cup [\![P_n]\!]_p) \cdot [\![P_1]\!]_p)^* && \text{(by definition of } [\![\cdot]\!]_p) \\
&= [\![P_1]\!]_p \cdot (([\![P_2]\!]_p \cup \cdots \cup [\![P_n]\!]_p \cup \tau) \cdot [\![P_1]\!]_p)^* && \text{(since } \tau \text{ is neutral)} \\
&= [\![\mathsf{p2t}(P_1)]\!]_t \cdot (([\![\mathsf{p2t}(P_2)]\!]_t \cup \cdots \cup [\![\mathsf{p2t}(P_n)]\!]_t \cup \tau) \cdot [\![\mathsf{p2t}(P_1)]\!]_t)^* && \text{(by induction)} \\
&= [\![\mathsf{p2t}(P_1)]\!]_t \cdot ([\![xor(\tau, \mathsf{p2t}(P_2), \ldots, \mathsf{p2t}(P_n))]\!]_t \cdot [\![\mathsf{p2t}(P_1)]\!]_t)^* && \text{(by definition of } [\![\cdot]\!]_t) \\
&= [\![\mathsf{p2t}(P_1)]\!]_t \cdot ([\![xor(\tau, \mathsf{p2t}(P_2), \ldots, \mathsf{p2t}(P_n))]\!]_t \cdot [\![\mathsf{p2t}(P_1)]\!]_t \cdot \tau)^* && \text{(since } \tau \text{ is neutral)} \\
&= [\![\mathsf{p2t}(P_1)]\!]_t \cdot [\![sand(\tau, \mathsf{p2t}(xor(P_2, \ldots, P_n)), \mathsf{p2t}(P_1))^r]\!]_t && \text{(by definition of } [\![\cdot]\!]_t) \\
&= [\![\mathsf{p2t}(P_1)]\!]_t \cdot [\![\mathsf{p2t}(\to (\tau, xor(\tau, P_2, \ldots, P_n), P_1)^r)]\!]_t \cdot \tau && \text{(by definition of } [\![\cdot]\!]_t) \\
&= [\![sand(\tau, \mathsf{p2t}(P_1), \mathsf{p2t}(\to (\tau, P_1, \to (\tau, xor(\tau, P_2, \ldots, P_n), P_1)^r))]\!]_t && \text{(by definition of } [\![\cdot]\!]_t) \\
&= [\![\mathsf{p2t}(\to (\tau, P_1, \to (\tau, xor(\tau, P_2, \ldots, P_n), P_1)^r))]\!]_t && \text{(by definition of } \mathsf{p2t}) \\
&= [\![\mathsf{p2t}(\circlearrowright (P_1, \ldots P_n))]\!]_t && \text{(by Lemma 1)}
\end{aligned}
$$

## C   Testing Correctness of our Tool

Experience shows that errors are easy to be introduced in the implementation of tools, even if based on a solid formal semantics, as in the case of our framework. We have conducted several tests aimed at checking the correctness of the implemented translator. The main set of experiments consists on verifying the equivalence between the input Process Tree and the resulting Attack Tree. The test consists of generating traces from Process Trees, and then verifying if the traces can be replayed by the translated Attack Trees. The experiment is conducted on a large scale, where we test a total of 1 000 Process Trees of different dimensions. The increase in the dimensions also allows us to verify the scalability of the approach. The procedure consists of the following 4 steps.

1. Firstly, we randomly generate the 1 000 Process Trees by means of a function from the PM4py library. To construct the set of Process Tree models, three different parameter configurations were adopted, reported in Table 5. In each run, we increased the size of the models to be generated. To set the size we make use of the mode, min and max parameters listed in the Table. These parameters are used to compute the triangular distribution which is used by the PM4py function to generate the list of activities. Furthermore, the probabilities of the operators reported in Table refer to the probability of observing the specific operator, i.e. a fair distribution in all the run.

2. Once the Process Trees are obtained, 1 000 traces are generated from each tree using a function from the same library.

3. The Process Trees are finally translated into Attack Trees by means of our translator. The tool also generates the corresponding RisQFlan file for each Attack Tree.

4. Finally, we verify if each trace generated by each Process Tree can be executed by the corresponding Attack Tree. For this step, we implemented an algorithm that, given a trace and an Attack Tree, verifies if the sequence of activities in the trace can be executed by the Attack Tree. Therefore, we use the depth-first search algorithm to verify the path.

As a result, for all the 1 000 Process Trees generated, the fitness value for the replay of all the 1 000 traces on the translated Attack Tree is 1. In other words, all traces generated by each Process Tree can be executed by the corresponding, translated, Attack Tree. This gives us high confidence in the correctness of the translator since we assessed the equivalence between an input Process Tree and the resulting Attack Tree.

The other aspect we evaluated is the processing time of the approach, calculated on the three runs mentioned above. We computed the processing time as the time to translate each Process Tree into Attack Tree, and save the Attack Tree file (i.e. step 3 above, without considering the conversion into RisQFlan file). The Process Tree was already loaded in the application as it was randomly generated by the mentioned function. The translation took on average 5.54, 14.62, 57.96 milliseconds for conf1, conf2 and conf3 respectively. The execution time picked at 22.14ms in the first configuration, 38ms in the second and 261.09ms in the third one. We noticed that the execution time increased with the increase of the size of the tree, but still remaining very efficient.

| Parameters | conf1 | conf2 | conf3 |
|---|---|---|---|
| No. of models | 300 | 300 | 400 |
| Mode activities | 30 | 50 | 150 |
| Min no. activities | 30 | 50 | 150 |
| Max no. activities | 50 | 100 | 300 |
| Prob. sequence op. | 0.25 | 0.25 | 0.25 |
| Prob. choice op. | 0.25 | 0.25 | 0.25 |
| Prob. parallel op. | 0.25 | 0.25 | 0.25 |
| Prob. or op. | 0.25 | 0.25 | 0.25 |

Table 5: Configurations of the three groups of Process Trees generated